\documentclass{iopart}
\usepackage{graphics}
\usepackage{epstopdf}
\usepackage{subfigure}

\def\ee{\end{equation}}

\def\mmax{\textmd{max}}

\def\rre{\textmd{Re}}
\def\iim{\textmd{Im}}

\def\iint{\textmd{int}}

\def\ttot{\textmd{tot}}

\def\ccor{\textmd{correl}}
\def\be{\begin{equation}}
\def\ee{\end{equation}}
\def\beq{\begin{equation}}
\def\eeq{\end{equation}}
\def\bea{\begin{eqnarray}}
\def\eea{\end{eqnarray}}

\def\beq{\begin{equation}}
\def\eeq{\end{equation}}
\def\bea{\begin{eqnarray}}
\def\eea{\end{eqnarray}}
\newcommand{\eins}{\mbox{$1 \hspace{-1.0mm} {\bf l}$}}
\usepackage{bm}
\usepackage{graphicx}

\begin{document}
\bibliographystyle{unsrt}
\title{Effects of dissipation in an adiabatic quantum search algorithm}
\author{In{\'e}s de Vega$^{1,2}$, Mari Carmen Ba{\~n}uls$^{2}$ and A. P\'erez$^{3}$}
\address{$^{1}$ Institut f\"ur Theoretische Physik, Albert-Einstein-Allee 11, Universit{\"a}t Ulm, D-89069 Ulm, Germany.}
\address{$^{2}$ Max-Planck-Institut f\"ur Quantenoptik,  Hans-Kopfermann-Str. 1, Garching, D-85748, Germany.}
\address{$^{3}$ Departament de F\'isica Te\'orica and IFIC, Universitat
  de Valencia-CSIC, Dr. Moliner 50, 46100-Burjassot, Spain.}
\ead{ines.devega@uni-ulm.de}
\begin{abstract}
We consider the effect  of two different environments on the
performance of the quantum adiabatic search algorithm, a thermal bath at finite
temperature, and a structured environment similar to the one
encountered in systems coupled to the electromagnetic field that
exists within a photonic crystal. While for all the parameter regimes
explored here, the algorithm performance is worsened by the contact
with a thermal environment, the picture appears to be different when
considering a structured environment. In this case we show that, by
tuning the environment parameters to certain regimes, the algorithm
performance can actually be improved with respect to the closed system
case. Additionally, the relevance of considering the dissipation rates
as complex quantities is discussed in both cases. More particularly,
we find that the imaginary part of the rates can not be neglected with the usual argument that it simply amounts to an energy shift, and
in fact influences crucially the system dynamics.
\end{abstract}
\maketitle
\section{Introduction}


The paradigm of Adiabatic Quantum Computation (AQC), introduced
in~\cite{farhi00}, is known to be equivalent to standard,
circuit based, quantum computation~\cite{aharonov04}.  From the point
of view of its physical implementation, it is however specially
appealing, since it is entirely formulated in terms of a controlled
dynamics, and moreover it offers some inherent robustness to
errors~\cite{childs01}.

In AQC the solution of a computational problem is encoded in the ground
state of a certain (problem) Hamiltonian. The system is started in
the easily preparable ground state of an initial Hamiltonian, which
is then slowly turned into the problem one. The adiabatic theorem~\cite{Born28,Kato1950,messiah,Ambainis2004,Lidar2009}
guarantees that, if the variation of the Hamiltonian is sufficiently
slow, time evolution will drive the system into the solution state.
The resource measuring the computational cost of the algorithm is
thus the total evolution time required to guarantee the adiabaticity
condition\footnote{A more strict measure of the cost is given by 
the product of the norm of the Hamiltonian times the total time~\cite{aharonov04}.}. 
For most cases, it depends on the inverse squared of the
minimum energy gap during the evolution.

Since quantum systems are not in general completely isolated from
their environment, errors and dissipation are ubiquitous, and it is
fundamental to decide how quantum computers and algorithms can be
built that achieve their computational tasks in spite of inaccurate
operations and certain loss of coherence.  But AQC gets affected by
errors in a fundamentally different way than the standard, gate-based
model of quantum computation. The interaction with the environment may
excite the system out of its ground state, causing errors, but if the
energy scales in the environment are much smaller than the minimum
gap, the adiabatic evolution will be naturally preserved. On the other
hand, the dissipation may also result in a modified effective
Hamiltonian, and thus affect the performance of the AQC \cite{lidar05,amin08,aminrole09}.  Even when
the coupling with the environment is weak, it may produce noticeable effects
in the performance of the adiabatic algorithm, specially considering
the large time scales that this requires.

Looking at the question from a different perspective, the algorithm may be implemented within a system inwhich the interaction with the environment is not an undesirable
effect, but rather a tool that can be actually controlled and tuned at
will to improve the performance of AQC.
Whether the effect of the environment is controllable or not, 
the system in question has to be considered as a \textit{quantum} open system, and its
interaction with the environment can be treated within the
\textit{weak coupling limit}, provided that this coupling is small
enough in comparison to the system and the environment time scales.


In this work we study the effect of different types of noise on a
particular adiabatic algorithm, namely the adiabatic version of Grover's search.  Grover's
problem, or that of search in an unstructured
database~\cite{grover97}, is one of the problems for which quantum
computation has been explicitly shown to exhibit a remarkable speedup
over the classical one. The adiabatic version of the quantum search algorithm
achieves, in a closed system, the optimal quadratic speedup with
respect to its best classical counterpart~\cite{roland02}.  But if the
system is subject to some dissipative dynamics, the performance of the
search may depend on the characteristics of the bath~\cite{tiersch07}.
In particular, it was pointed out by Amin and
collaborators~\cite{amin08} that the presence of a thermal bath can in
some cases enhance the performance of the adiabatic quantum search.



In this paper, we study the performance of the adiabatic search algorithm implemented
in a quantum open system. To this end, we adopt the two-level approximation,
in which only the interaction of the environment with the two levels 
involved in the quantum search algorithm is considered~\cite{aminrole09,tiersch07}.

We will consider two situations.

First, we analyze the case in which the system is coupled to a thermal
environment. In particular, we consider an ohmic environment as
appearing in~\cite{LCD87}, which is valid to describe cases in
which the spectral density is a smooth function in the frequency
space.

Second, we consider an ideal situation in which the adiabatic search
algorithm is implemented within an atom lattice. In that case, we
study the model proposed in~\cite{deVega08}, where the atoms in the
lattice are coupled in a controlled way to a bosonic environment which
has very similar properties to the radiation field within a photonic
crystal. Here we study how, under certain conditions, and always
within the two-level approximation, a controlled coupling with the
environment can give rise to an improvement of the quantum search
algorithm.  To study the generality of this result, we analyze a real
two-level system (as opposed to an effective one arising from the
interplay of many qubits) which undergoes an adiabatic evolution,  
not necessarily corresponding to the search algorithm, and which interacts
with the same photonic-crystal-like environment.  We show that, even in
this situation, the coupling with the environment makes the system, in
certain parameter ranges, end up in a final state that corresponds,
with a higher probability, to the \textit{target state} (the ground
state of the final Hamiltonian).

In both cases, the quantum open system is 
described through a master equation formalism. Hence, the effect of
the environment in the system dynamics is encoded in a collection of
terms that depend on the so called dissipation rates. In this paper we
stress the importance of considering these rates as complex
quantities. 
Indeed, the conclusions about the performance of the
adiabatic algorithm highly depend on whether the imaginary parts are
neglected or not, which suggests that they cannot be considered simply
as an overall energy shift.  We use units in which $\hbar=1$ .

\section{Adiabatic Grover's algorithm}

In Grover's problem, the goal is to find a particular item, $m$, in an
unstructured database of size $N=2^{n}$.  
The best classical algorithm for
this search takes time $O(N)$, while it was shown in~\cite{grover97}
that there is a quantum algorithm solving the problem in
time $O(\sqrt{N})$, known to be the optimal performance~\cite{bennett97,zalka99}.

The quantum algorithm maps each element of the database onto one
element of the computational basis for $n$ qubits.  
The solution will then correspond to a particular state, $|m\rangle$.  In the
adiabatic algorithm~\cite{roland02}, the system is started in an easy
to prepare initial state, containing an equal superposition of all
basis states, 
\bea
|\Psi_0\rangle=\left [\frac{1}{\sqrt{2}}\left(|0\rangle+|1\rangle\right)\right ]^{\otimes n}=\frac{1}{\sqrt{N}}\sum_{z=0}^{N-1}|z\rangle,
\eea 
where $|z\rangle$ are the elements of the computational basis.
Then a time dependent Hamiltonian is applied which smoothly
interpolates between the initial Hamiltonian \beq
H_0=\eins-|\Psi_0\rangle \langle \Psi_0|, \eeq having
$|\Psi_0\rangle$ as its ground state, and the final one, \beq
H_m=\eins-|m\rangle\langle m|, \eeq whose ground state is the
solution $|m\rangle$.  
If the adiabatic condition is satisfied, the final state will be 
$|\Psi_f\rangle\approx |m\rangle$.

The Hamiltonian governing the evolution can be written as
\beq
H_G(s)=(1-s) H_0 + s H_m,
\label{Grov1}
\eeq
in terms of the dimensionless parameter $s\in[0, 1]$, a function of 
time that controls how fast the Hamiltonian changes.
$H_G$ acts non-trivially only on the 
two dimensional subspace spanned by $|m\rangle$ and 
$|m_{\perp}\rangle=\frac{1}{\sqrt{N-1}}\sum_{z\neq m}|z\rangle$,
and can thus be diagonalized analytically.
Indeed, in this subspace it can be expressed as
\bea
H_G(s)=\frac{1}{2}\eins+H(s),
\label{Grov2}
\eea
where we have defined 
$H(s)=\frac{1}{2}(\Omega(s)\sigma_x-\Delta(s)\sigma_z)$ with $\Delta(s)=2\frac{1-s}{N}+(2s-1)$ and $\Omega(s)=2(s-1)\frac{\sqrt{N-1}}{N}$, 
$\sigma_{x,z}$ being the Pauli matrices in the basis 
$\{|m\rangle ,\ |m_{\perp}\rangle \}$. 
The time-dependent energy eigenvalues are 
$E_{0,1}(s)=\frac{1}{2}\mp \frac{\sqrt{\Omega(s)^2+\Delta(s)^2}}{2}$,
and the corresponding eigenvectors
\bea
|0(t)\rangle&=&\sin\theta |m\rangle+ \cos\theta|m_{\perp}\rangle,\nonumber\\
|1(t)\rangle&=&-\cos\theta|m\rangle+\sin\theta |m_{\perp}\rangle,
\eea
where $\sin\theta=\sqrt{\frac{\Delta(s)+\alpha(s)}{2\alpha(s)}}$
and $\cos\theta=-\frac{\Omega(s)}{\sqrt{2 \alpha(s) (\alpha(s)+\Delta(s))}}$.
The subspace orthogonal to $|m\rangle$ and $|m_{\perp}\rangle$
is the eigenspace of $H_G$ with energy $E_2=1$.
The time dependent gap between the ground and first excited state 
is then given by $\alpha(s)=E_1-E_0=\sqrt{\Omega^2+\Delta^2}$.

The adiabatic condition imposes a lower bound on the 
running time of the algorithm, 
$T\gg \max \frac{\langle 1(s)|\frac{d H}{d s}|0(s)\rangle}{\alpha(t)^2}$, 
where the maximum is taken over 
the whole range $s\in[0,1]$.

If the interpolation is linear, i.e. $s_{lin}(t)=t/T$, 
and the total time is larger than $T_{lin}=\frac{N}{\epsilon}$,
the adiabatic condition is approximately satisfied, 
with an error $\epsilon^2$ in the overlap between 
the final state and the solution. 
Using a different function $s(t)$, it is possible to ensure the condition
by adjusting the velocity of change of the Hamiltonian to the instantaneous
gap~\cite{roland02}. This leads to the optimal performance, when
the interpolating function is \beq s_{opt}(t)= \frac{1}{2}
\left ( 1+ \frac{\tan( 2\epsilon t \frac{\sqrt{N-1}}{N}-
\arctan{\sqrt{N-1}} )}{\sqrt{N-1}} \right ). \eeq In
that case, the total running time of the algorithm is $T_{opt}\simeq\pi\sqrt{N}/2\epsilon$
for $N\gg1$. 


\section{Evolution equation of the system weakly coupled to an environment}
Let us consider the evolution of a system evolving adiabatically with
Hamiltonian $H_S(t)$, and interacting with an environment whose free
Hamiltonian is $H_B$. Then, the total Hamiltonian has the form
\be
H=H_S(t)+H_B+H_\iint,
\label{total}
\ee 
where $H_\iint$ is the interaction Hamiltonian, which describes a
linear coupling between system and environment operators. In
particular, for an environment given as a set of harmonic
oscillators, described by annihilation $b_\lambda$ and creation
$b^\dagger_\lambda$ operators, the interaction may have
the form 
\bea 
H_\iint=\sum_\lambda\sum_i g^i_\lambda(\sigma_i
b_k^\dagger+\textmd{h.c.})
\label{nqubits}
\eea 
where $g^i_\lambda$ is the coupling constant of the system with
the mode $\lambda$, and $\sigma_i$ ($\sigma_i^\dagger$) are spin
ladder operators corresponding to the qubit $i$.

In our case, we will consider that the environment is mainly coupled
to the two-dimensional subspace spanned by $|m\rangle$ and $|m_{\perp}\rangle$~\cite{tiersch07,aminrole09}.
In this situation, the
interaction Hamiltonian (\ref{nqubits}) can be written approximately~\cite{perez09}
as $H_{\iint}=A\otimes B$, where $A=\sigma_z$ and $B$ are operators
that act on the system and the environment Hilbert spaces, respectively.
Considering that the environment is composed of a set of harmonic
oscillators, the environment coupling operator can be written as $B=\sum_{\lambda}g_{\lambda}(b_{\lambda}^{\dagger}+b_{\lambda})$.
The evolution equation of the system within the weak coupling approximation
is given by (see the appendix for further details),
\bea
\frac{d\rho_S}{dt}&=&-i[H_S(t),\rho_S]-\int_0^t d\tau g(t-\tau)\left(AA(\tau,-t)\rho_S-A(\tau,-t)\rho_S A\right)\nonumber\\
&-&\int_0^t d\tau g^*(t-\tau)\left(\rho_S A(\tau,-t)A-A\rho_S A(\tau,-t)\right).
\label{red2}
\eea
with $A(\tau,-t)={\mathcal U}(t){\mathcal U}^\dagger(\tau)A {\mathcal U}(\tau){\mathcal U}^\dagger(t)$, ${\mathcal U}(t)={\mathcal T} \exp [-i \int_0^t H_S(\tau) d\tau]$ (${\mathcal T}$ being the usual time-ordering operator), and 
\begin{eqnarray}
g(t)= \sum_\lambda g_\lambda^2 [\coth{\left(\frac{\beta\omega_\lambda}{2}\right)}\cos{(\omega_\lambda t)}-i\sin{(\omega_\lambda t)}].
\label{correlh}
\end{eqnarray}
The well known Lindblad equation can be recaptured from the former equation by just considering a delta-correlated bath, so that $g(t-\tau)=\Gamma\delta(t-\tau)$. In such a case, $\int_0^t d\tau A(\tau,-t)g(t-\tau)=\Gamma A$, and similarly $\int_0^t d\tau A(\tau,-t)g^*(t-\tau)=\Gamma^* A$.
Hence, the Lindblad equation is not valid for every system that is
weakly coupled with its environment. It requires, in addition, that
the coupling operators of the system (in our case there is a single
one, $A$) evolve very slowly with the system Hamiltonian, within the
time scale $\tau_C$ in which the correlation function $g(\tau)$
decays.



\subsection{Bloch-Redfield equation}
The Bloch-Redfield equation describes the evolution of the density operator in the diagonal basis $|n\rangle$ of $H_S$, $\dot{\rho}_{nm}=\langle m|\dot{\rho}_S|n\rangle$, with $\rho_{mn}=\langle m|\rho_S|n\rangle$. In our case, the system Hamiltonian is time dependent, and the time dependent eigenbasis $|n(t)\rangle$ (corresponding to the set of eigenvalues $E_n(t)$) that diagonalizes instantaneously the system Hamiltonian $H_S(t)$, should be considered. In this situation, the matrix elements of the system density operator are $\rho_{mn}=\langle m (t)|\rho_S|n (t)\rangle$, and their evolution equation can be written as
\bea
\dot{\rho}_{nm}=\langle m (t)|\dot{\rho}_S|n (t)\rangle+\langle\dot{m}(t)|\rho_S|n(t)\rangle+\langle m(t)|\rho_S|\dot{n}(t)\rangle.
\label{bred1}
\eea
The first term on the right hand side of the equation is just the usual master equation (\ref{red2}), expressed in the instantaneous basis of $H_S(t)$.
In order to project (\ref{red2}) on the system basis, one should calculate terms such as $A(\tau,-t)|n(t)\rangle$, which requires the calculation of quantities such as $\mathcal{U}(t)|n(t)\rangle$.
Considering as a good approximation that the free system is undergoing
an exact adiabatic evolution, so that the adiabatic theorem can be
applied, we can write \cite{Mostafazadeh97}
\begin{eqnarray}
\mathcal{U}(t)|n(t)\rangle=e^{-i \int_0^t E_n(\tau)d\tau}|n(t)\rangle.
\label{adiav1}
\end{eqnarray}
Hence the first term of the evolution equation (\ref{bred1}) can be expressed as
\begin{eqnarray}
\langle n (t)|\dot{\rho}_S|p (t)\rangle&=&-iE_{np}\rho_{np}+\sum_{ql}\Gamma_{qn}(t)A_{nq}\rho_{ql}A_{lp}
+\sum_{ql}\Gamma^*_{pq}A_{nl}\rho_{lq}A_{qp}\nonumber\\
&-&\sum_{lq}\Gamma_{ql}A_{nl}A_{lq}\rho_{qp}-\sum_{ql}\Gamma^*_{pq}A_{nl}\rho_{lq}A_{qp},
\end{eqnarray}
where $A_{nm}=\langle n(t)|A|m(t)\rangle$, $E_{np}=E_n-E_p$, and
\bea
\Gamma_{ql}(t)=\int_0^t d\tau g(t-\tau)e^{i\int_\tau^t d\tau' (E_q(\tau')-E_l(\tau'))}
\label{gammar}
\eea 
are the dissipation rates mentioned in the introduction.  Notice
that in our two level system, any energy difference is
$E_q(t)-E_l(t)=\pm \alpha(t)$, where $q,l=0,1$. 
In order to further simplify the equations
we may assume that the rate of variation of $\alpha(t)$
is much slower than the decaying of the correlation function $g(t)$.
In other words, when the condition
\begin{eqnarray}
\frac{1}{\alpha(t)}\frac{d\alpha}{ dt}\ll \frac{1}{\tau_C}
\end{eqnarray}
is satisfied, the integrand $E_q(\tau')-E_l(\tau')=\pm \alpha(\tau')$ appearing in (\ref{gammar}) will variate very slowly in the integration region, and we can write
\bea
\Gamma_{ql}(t)=\int_0^t d\tau g(t-\tau)e^{i(E_q(t)-E_l(t))(t-\tau)}.
\label{gamma1}
\eea
 
Let us consider the evolution equations for our case, where the coupling operator $A=\sigma_z$.
Expressing the density operator $\rho_S$ in the spin basis $\{\sigma_x(t),\sigma_y(t),\sigma_z(t)\}$ as $\rho_S=\frac{1}{2}(1+\rho_x\sigma_x(t)+\rho_y\sigma_y(t)+\rho_z\sigma_z(t))$, the master equation can be written as follows
\bea
\dot{\rho}_x&=&2\textmd{cs} \Gamma^-_R-4\textmd{c}^2\rre(\Gamma_{00})\rho_x
+\alpha(t)\rho_y+2\bigg(\textmd{cs}\Gamma^+_R
+\dot{\theta}\bigg)\rho_x
\nonumber\\&+&2\textmd{cs}\bigg(\Gamma^+_R-2\rre(\Gamma_{00})\bigg)\rho_z\nonumber\\
\dot{\rho}_y&=&2\textmd{cs} \bigg(\Gamma_I^+ -2\iim(\Gamma_{00})\bigg)+\bigg(2\textmd{s}^2\Gamma_I^+-\alpha(t)\bigg)\rho_x
\nonumber\\
&-&2\bigg(2\textmd{c}^2\rre(\Gamma_{00})+\textmd{s}^2
\Gamma_R^+\bigg)\rho_y+2\textmd{cs}\Gamma_I^- \rho_z\nonumber\\
\dot{\rho}_z&=&2\textmd{s}^2 \Gamma_R^- -2(2\textmd{cs}\rre(\Gamma_{00})-\dot{\theta})
+2\textmd{s}^2\rre(\Gamma_{00}))\rho_x+2\textmd{cs}(\Gamma_R^+
\nonumber\\
&-&2\rre(\Gamma_{00}))\rho_z,
\label{evol}
\eea
where $\Gamma^-_R=\rre(\Gamma_{10}-\Gamma_{01})$, $\Gamma^+_R=\rre(\Gamma_{10}+\Gamma_{01})$, $\Gamma^-_I=\iim(\Gamma_{10}-\Gamma_{01})$ and $\Gamma^+_I=\rre(\Gamma_{10}+\Gamma_{01})$. The dissipation rates are defined as
\bea
\Gamma_{00}=\int_0^t d\tau g(\tau),\nonumber\\
\Gamma_{01}=\int_0^t d\tau g(\tau) e^{i\alpha(t)\tau},\nonumber\\
\Gamma_{00}=\int_0^t d\tau g(\tau)e^{-i\alpha(t)\tau},
\label{rates0}
\eea
and the variables
$\textmd{c}=\cos{2\theta}=-\frac{\Delta}{\alpha}$, and $\textmd{s}=\sin{2\theta}=-\frac{\Omega}{\alpha}$.

\section{Adiabatic evolution in different environments}

We now analyze two different cases in which adiabatic evolution occurs
in the presence of dissipation.

In the first place we consider the effect of a thermal environment on
the performance of the adiabatic search algorithm. A thermal environment is the
one that we would expect to find naturally when a system is coupled to
a bosonic bath, for instance phonons in a solid lattice, or the
radiation field at finite temperature.

In the second case we consider our system to be coupled in a
controlled way to a bosonic environment that corresponds to a matter
wave field. This example is much more specific than the former, but
allows us to illustrate a case in which, provided that the two-level
approximation remains valid, an artificially controlled coupling
allows in some regimes an improvement on the performance of an adiabatic
quantum computation.

In all our simulations we consider the number of qubits $n=10$.

\subsection{Thermal environment}
\label{thermal}

As noted above, the most
important quantity that characterizes the influence of an environment
on the system dynamics is the so-called correlation function. When the
exact form of the coupling constants $g_k$ or the dispersion relation
of the environment $\omega({\bf k})$ is not known, a phenomenological
model should be used to describe the interaction. In that situation,
we can express the correlation function (\ref{correlh}) as
\begin{eqnarray}
g(t)=\int_0^\infty d\omega J(\omega)[\coth{\left(\frac{\beta\omega}{2}\right)}\cos{(\omega t)}-i\sin{(\omega t)}],
\label{spectral}
\end{eqnarray}
which fulfills the property $g(-\tau)=g^*(\tau)$. In the last
expression, the function $J(\omega)$ is the so-called spectral density
of the bath.  A very well known approach consists in 
assuming that $J(\omega)$ behaves as $\omega^s$~\cite{LCD87}. In that
case, it can be written as
\begin{eqnarray}
J(\omega)=\eta \omega^s \omega^{1-s}_c e^{-\omega/\omega_c}.
\label{chapuno41}
\end{eqnarray}
This model of spectral function has been extensively studied in the
context of the spin-boson model \cite{LCD87}, where three different
regimes were described: a \textit{sub-ohmic} regime in which $0<s<1$,
a \textit{super-ohmic} regime with $s>1$, and an \textit{ohmic} regime
where $s=1$.
The exponential factor appearing in the last expression has been added
to provide a smooth cut-off for the spectral density, which is
modulated by the frequency $\omega_c$. This parameter controls the
correlation time of the environment, approximately given by
$\tau_c\sim 1/\omega_c$: the larger $\omega_c$, the smoother the
spectral function, and the shorter the time the environment
takes to relax to equilibrium, giving rise to a more Markovian
interaction.  As seen in \cite{LCD87}, whether we have \textit{ohmic},
\textit{sub-ohmic} or a \textit{super-ohmic} spectral function depends
on the type of reservoir, and determines quite strongly the evolution
behavior of the coupled system. In this work we will focus on one of
the most significant cases, the ohmic dissipation, in which
$s=1$, and on large $\omega_c$, when the spectral function is a smooth
function of the frequency.

\begin{figure}[floatfix]
\begin{minipage}[c]{.4\columnwidth}
\subfigure[Fig 1.a.]{
 \label{figure1A}
 \includegraphics[width=\columnwidth]{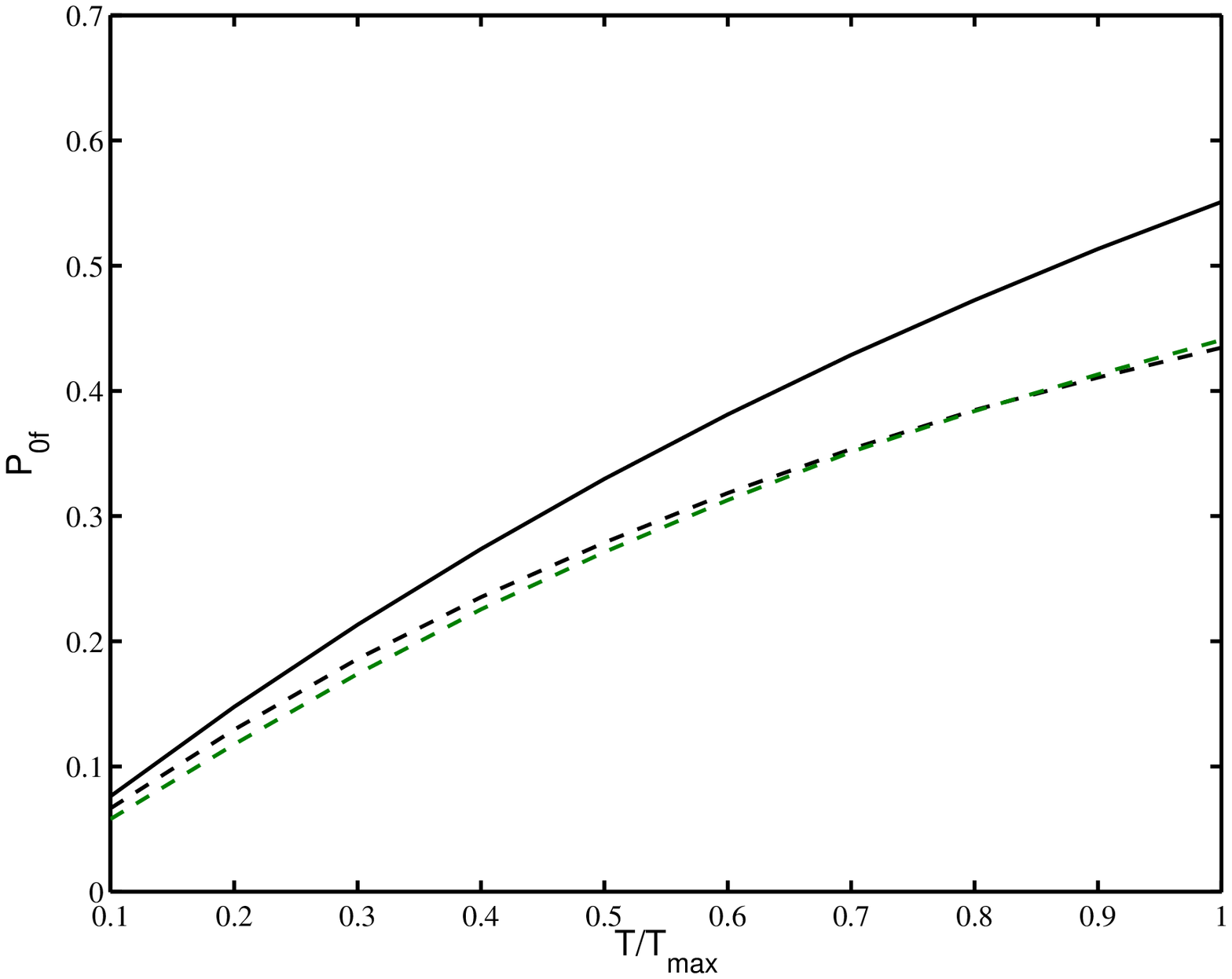}
}
\end{minipage}
\hspace{.05\columnwidth} 
\begin{minipage}[c]{.4\columnwidth}
\subfigure[Fig. 1.b.]{
 \label{figure1B}
 \includegraphics[width=\columnwidth]{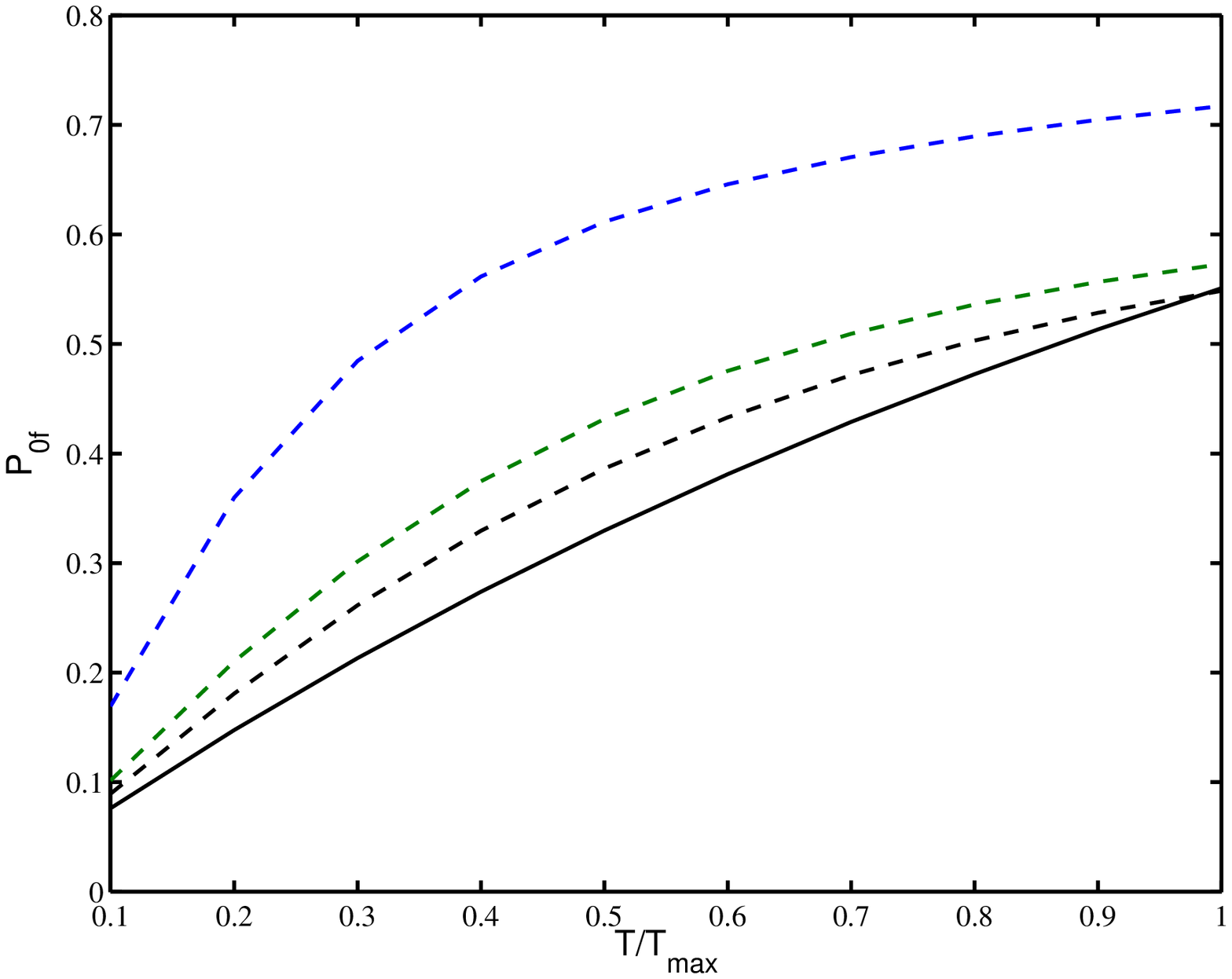}
}
\end{minipage}
\caption{Considering the coupling with a thermal environment, final success probability of the algorithm with adiabatic final adiabatic time $T$, as a function of $T/T_{\textmd {max}}$ (being $T_{\mmax}=0.8 T_{lin}$, with $T_{lin}=\frac{4N}{\pi}$, the time required for final success probability $0.55$ in the closed case) for different coupling parameters $\eta$, and for two cases: Fig 1.a. considering complex dissipation rates $\Gamma_{ij}$ (with $i,j=0,1$), as they appear naturally in the derived Bloch-Redfield equations, and Fig 1.b. considering only the real part of the dissipation rates $\Re\{\Gamma_{ij}\}$ (discontinuous lines).  In both figures, the black solid line corresponds to the closed case $\eta=0$, while black, green and blue discontinuous lines corresponds to $\eta=0.05,0.1,0.5$ respectively. The curve $\eta=0.5$ is not shown in Fig 1.a., since for that value the imaginary part of the rates is too large and gives rise to unphysical results, what points out a failure of the weak coupling assumption. The frequency cut is chosen as $\omega_c=0.25$.}
\end{figure}

Let us study how the final success probability of the adiabatic search algorithm
is affected by the presence of a thermal environment. To this order,
we consider the evolution equations derived in the former section,
with dissipation rates that depend on the correlation function
(\ref{spectral}). As shown in figure~\ref{figure1A}, the final
success probability decreases in the presence of dissipation for any value of 
evolution time of the adiabatic algorithm $T$ considered.  A common
approximation in the literature 
consists in neglecting the effect of the imaginary parts of the
dissipation rates (\ref{rates0}), with the argument that they amount
only to an energy shift which in the context of quantum optics is
known as the Lamb shift~\cite{quantumoptics}.  However, in the current
scenario, figure~\ref{figure1B} shows that if only the real part of
the dissipation rates is considered in the equations, the result for
the same couplings is completely different. This shows that, at
least in the present case, the imaginary parts of the rates cannot be
considered as a simple energy shift. In fact, when eliminating the
imaginary parts, one can see that the larger the coupling with the
environment, the higher the final success probability at any adiabatic
evolution time.
 
As a consequence, we stress that in order to correctly describe the
system, the imaginary parts of the rates should in general be
considered, since they appear naturally in the derivation of the
second order evolution equation of the system (in other words, they
correspond to second order terms, as well as the real parts), and they
produce relevant changes in the evolution.


\subsection{Structured environment}
\label{structured}

The effect of dissipation on the performance of the adiabatic search
will depend on the details of the environment and its interaction with
the system.  The particular features of the adiabatic search make it
specially sensitive to energy scales in the bath that are comparable
to the characteristic gap in the system, $\Delta
E$~\cite{childs01,amin08}.  Intuitively, one would expect that the
algorithm is more resilient when this energy corresponds to a gap in
the spectral function of the environment (i.e. a frequency region where this function is zero), similar to the effect when
the dominant energy scale of the bath is far away from $\Delta E$, as
studied for the thermal case in~\cite{amin08}.  
However, while this picture gives a qualitative image, a complete idea of the effect of the environment in the algorithm can only be achieved with a more complete study. 
With this aim, in the following we analyse in detail the effect of an structured environment in the performance of the search algorithm. Particularly, we will consider an environment with a gap in the density of states (which leads to a gap in the spectral density), and which has the same characteristics as the radiation field within a photonic crystal.
In addition to that, it is specially interesting to consider the case of a controllable environment, where
the capability to tune the parameters of the interaction may give us the
possibility to improve the performance of the algorithm 
via the dissipative dynamics. 
Such an environment is indeed realizable in the
context of optical lattices, as we discuss in the following.

Let us therefore consider an ideal situation
in which the quantum search algorithm is implemented in an atom
lattice, which is formed by loading a gas of ultracold atoms in a
standing wave field. In the last few years, atom lattices have been
experimentally realized by several groups
\cite{J98,Mandel03,G02,Foelling05,PC04,Bloch2}, and due to their high
controllability, they have also been proposed as a candidate to
realize a quantum computer \cite{Bloch2}.

In addition, it has been recently shown \cite{RFZDZ04,deVega08} that
atoms in an optical lattice can be coupled to an environment in a
controlled way. In the later proposal, atoms are considered to have
two relevant internal states $a$ and $b$. Atoms in state $a$ are
actually trapped by the optical lattice, and have a frequency
$\omega_a=\omega_a^0+\omega_0/2$, with $\omega_0$ the trap frequency
of the lattice. In addition, they are considered to be in the
so-called strongly correlated regime, where either there is one or
zero atoms at each site of the lattice. Atoms in state $b$ are not
trapped by the lattice potential, and have an energy
$\hbar\omega_b+\frac{\hbar k^2}{2m}$, where $\hbar\omega_b$ and
$\frac{\hbar k^2}{2m}$ correspond to the internal and kinetic energies,
respectively.  If a Raman transition of total frequency $\omega_L$ and
Rabi frequency $\Omega_L$ is produced between the trapped and the
untrapped state, the Hamiltonian that describes this process can be
written in the interaction picture as
\begin{eqnarray}
H_{\iint}&=&\sum_{\bf j} \sum_{\bf k}g_{k} \left(b^\dagger_{\bf
k}\sigma_{\bf j} e^{i \Delta_{\bf k} t-i({\bf k}-{\bf k}_L)\cdot{\bf
r}_{\bf j}}+h.c.\right). \label{genH2}
\end{eqnarray}
In (\ref{genH2}), the sum in $j$ runs over the $N$ sites in the lattice, ${\bf r}_{\bf j}$ denotes the positions in the lattice, and
$\Delta_{\bf k}=k^2/2m -\Delta_L$, with $\Delta_L=\omega_L-(\omega_b
-\omega_a)$ the laser detuning. The coupling constants are $g_k= \Omega_L e^{-X_0^2 k^2/2} (8\pi^{3/2} X_0^3/V)^{1/2}$, where
$X_0=(1/2m\omega_0)^{1/2}$ is the size of the wave function at each
site.

The previous Hamiltonian is very similar to the one describing the
interaction of two level atoms with the radiation field. However, here
the spin operators $\sigma_i$ (equivalently $\sigma_i^\dagger$) are
not describing transitions between atomic internal states, but rather
they describe transitions from a Fock state $|1\rangle_i$ (which
corresponds to the presence of an atom at site $i$), to the Fock state
$|0\rangle_i$, describing the absence of atoms at site $i$. On the
other hand, the operators $b^\dagger_{\bf k}$ (equivalently $b_{\bf
  k}$) correspond to the creation (annihilation) operators of a bath
of harmonic oscillators, which in this case is not the radiation
field, but the matter-wave field that describes the untrapped atoms.

In our model, we assume that the qubits undergoing adiabatic evolution
are encoded in the presence $|1\rangle_i$ or absence $|0\rangle_i$ of
an atom at site $i$ of the lattice.  The coupling with the environment
produces transitions between these basis states through the
Hamiltonian (\ref{genH2}). However, contrary to the case of the
thermal environment studied in the former section, here the coupling
with an environment is not an uncontrolled event, but rather it is artificially
produced through the two-photon Raman transition to the untrapped
level. For that reason, several external parameters that describe the
interaction can be easily controlled: most importantly $\Omega_L$, which
determines the coupling strength, $\Delta_L$, which determines the
resonance condition, and $\omega_0$, which determines the trapping
frequency of each lattice well, and as we will later see, will
also characterize the width of the spectral function in the frequency space
$J(\omega)$.

It is important to notice that, contrary to the thermal case, in this
situation the correlation function can be fully determined by the
coupling parameters $g_k$ and the dispersion relation of the
environment, given by $\omega_k=\frac{k^2}{2m}+\omega_b$. No
phenomenological model is needed to characterize the spectral
function. On the other hand, the relation $\omega_k$ indeed resembles that of
the radiation field in a three dimensional and infinite photonic
crystal near the band-gap edge, which here corresponds to the
frequency $\omega_b$ \cite{WJ03,JQ95,deVega05}. This
environment will give rise to a very particular dissipation in our
system. The correlation function can be written as \cite{deVega08}
\bea 
g(t)=\sum_{\bf k} g_k^2 e^{-i\Delta_{\bf k}t}=
 \Omega_L^2\frac{e^{i \Delta_L t}}{\nu_t^{3}},
 \label{Gn0}
 \eea 
where we have assumed
 that the sum in the wave vector $\bf{k}$ can be performed in the
 continuum limit. Here, the quantity $\nu_t=\sqrt{1+i\omega_0 t}$.

\begin{figure}[floatfix]
\begin{minipage}[c]{.4\columnwidth}
\subfigure[]{
 \label{PBG1}
 \includegraphics[height=.75\columnwidth]{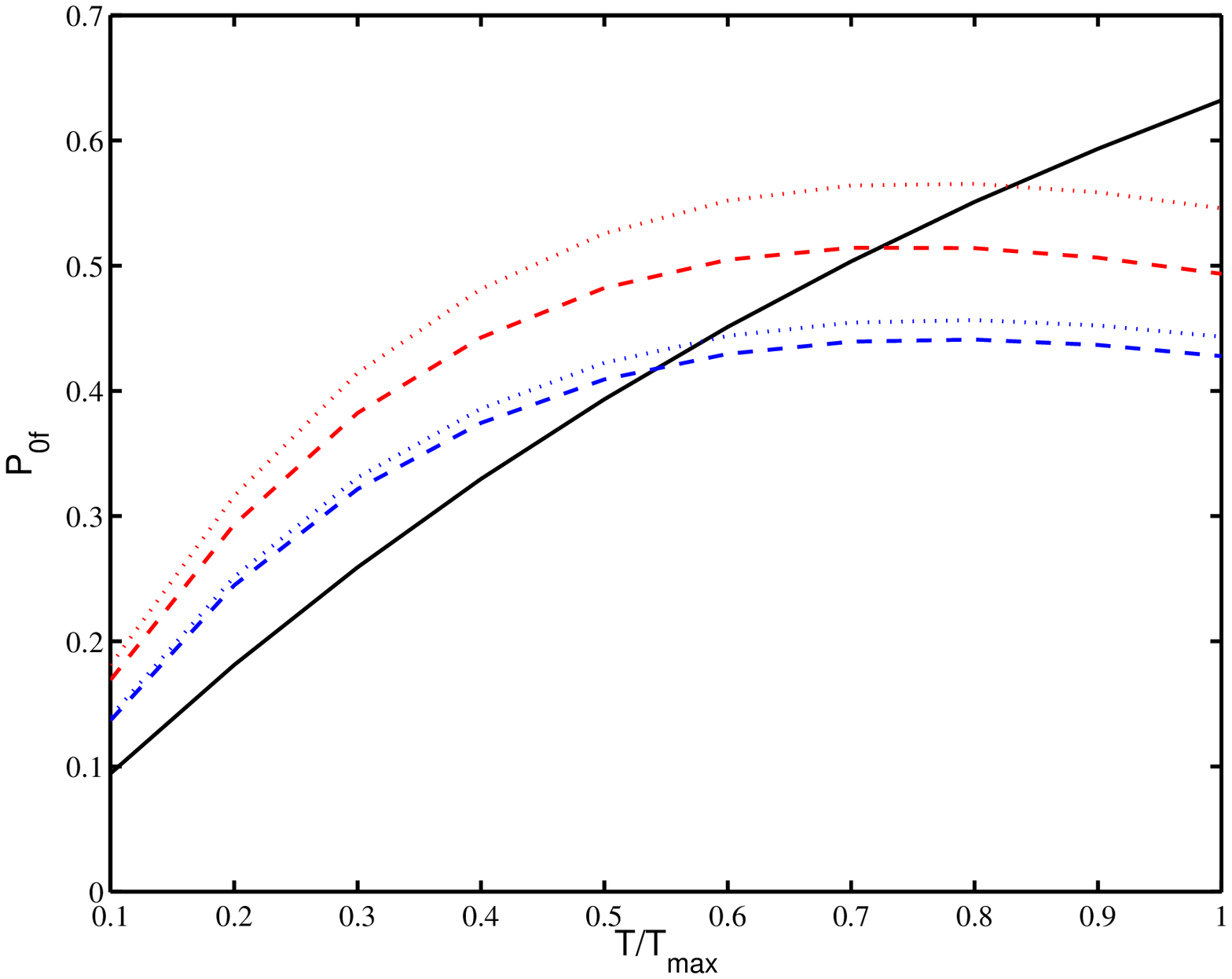}
}
\end{minipage}
\hspace{.050\columnwidth} 
\begin{minipage}[c]{.4\columnwidth}
\subfigure[]{
 \label{PBG2}
 \includegraphics[height=.75\columnwidth]{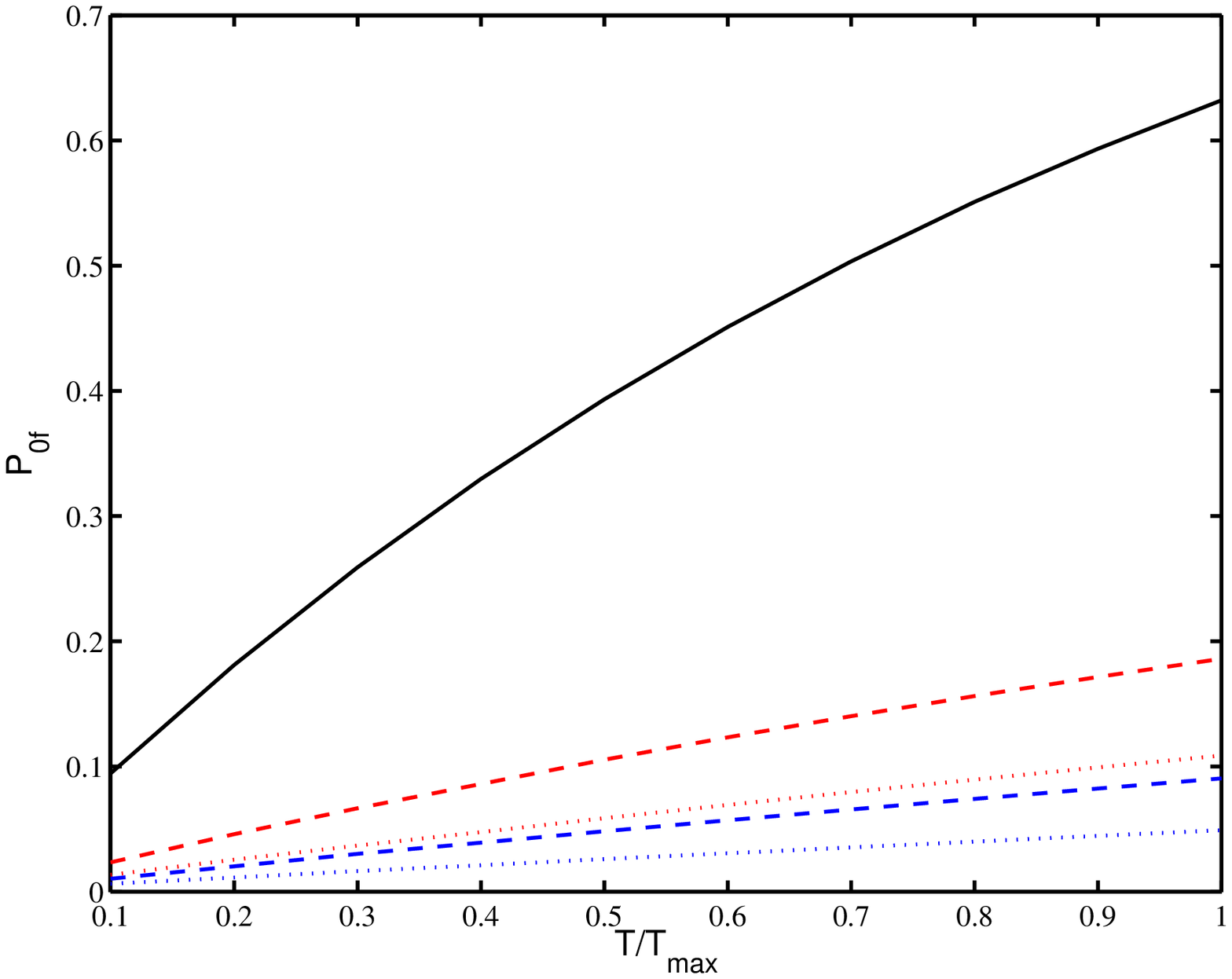}
}
\end{minipage}
\caption{For a system coupled with a structured environment, the
  figures show the final success probability of the algorithm as a
  function of $T/T_{\mmax}$ (with $T_{\mmax}=T_{lin}$, and
  $T_{lin}=\frac{4N}{\pi}$). Solid black lines represent the solution
  for a closed system; red and blue dashed lines correspond to
  $\eta=0.05$ with $\Delta_L=0.2$ and $0.27$ respectively; red and
  blue dotted lines correspond to $\eta=0.1$ with $\Delta_L=0.2$ and
  $0.27$ respectively. Fig 1.a. represents the curves for complex
  rates, and for different couplings $\gamma$ and laser detunings
  $\Delta_L$. Fig 2.a. represents the curves for the same parameters
  but considering only the real part of the dissipation rates. In both
  cases the trap frequency is chosen as $\omega_0=0.25$.}
\end{figure}

\begin{figure}[floatfix]
\begin{minipage}[c]{.4\columnwidth}
\subfigure[]{
 \label{PBG3}
 \includegraphics[height=.75\columnwidth]{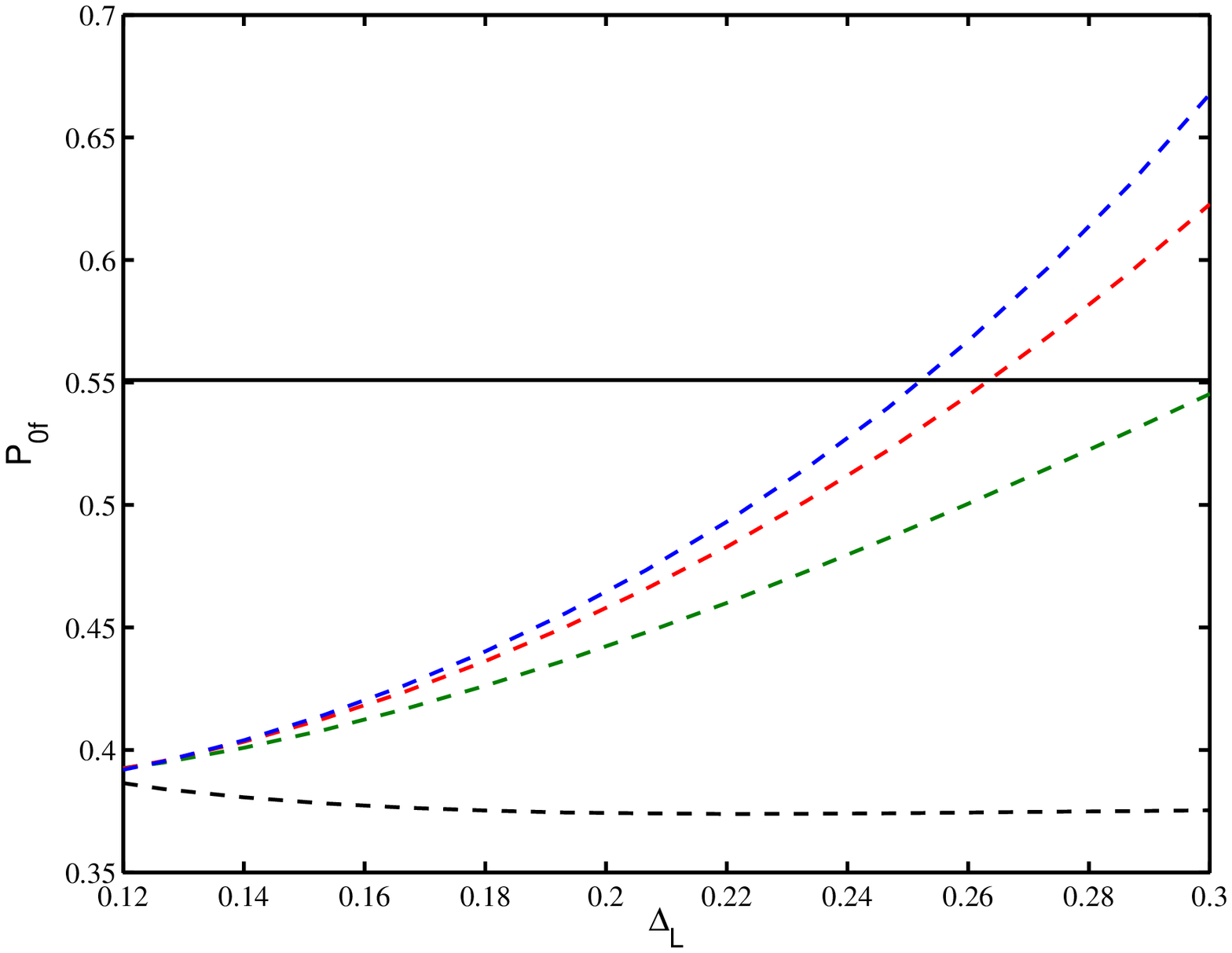}
}
\end{minipage}
\hspace{0.050\columnwidth} 
\begin{minipage}[c]{.4\columnwidth}
\subfigure[]{
 \label{PBG4}
 \includegraphics[height=.75\columnwidth]{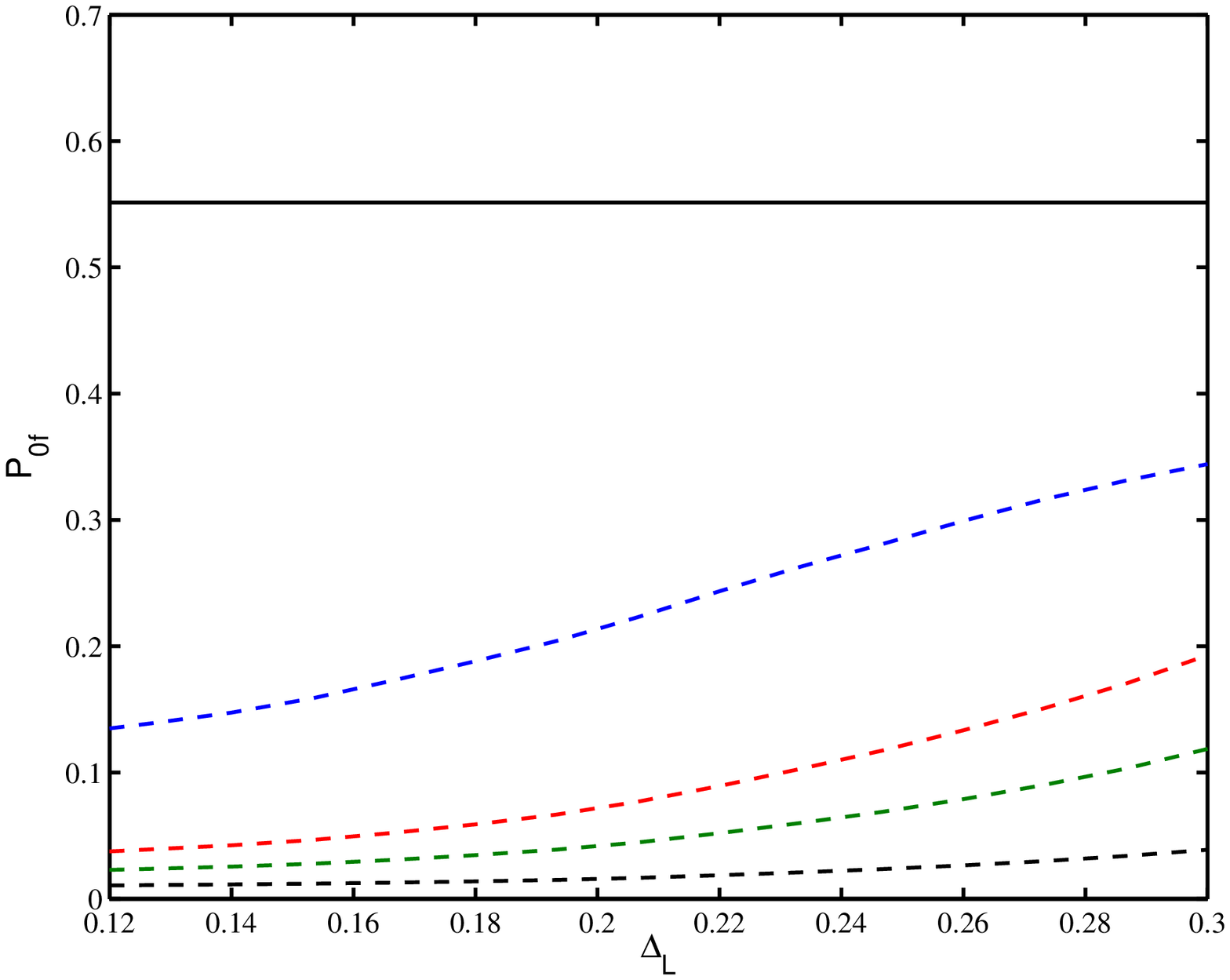}
}
\end{minipage}
\caption{For a system coupled with a structured environment, the final
  success probability of the algorithm is plotted as a function of the
  laser detuning $\Delta_L$. Fig. 3.a represents the solutions
  considering complex rates. In this figure, the solid black line
  represents the solution for the closed system, while black, green,
  red and blue discontinuous lines correspond to couplings
  $\eta=0.01,0.05,0.1,0.4$ (from the bottom up)
  respectively. Fig. 3.b. shows the curves considering only the real
  part of the dissipation rates. In this figure, the solid black line
  represents the solution for the closed system, while black, green,
  blue discontinuous lines correspond to couplings
  $\eta=0.01,0.05,0.2$ (from the bottom up) respectively. The trap
  frequency is chosen as $\omega_0=0.25$, and $T_{max}=0.8\ T_{lin}$,
  with $T_{lin}=\frac{4N}{\pi}$.}
\end{figure}

In figure~\ref{PBG1}  we observe, always in the
two-level approximation and for various parameter regimes, a larger
final probability of success at small values of $T$ for the open system
(discontinuous lines) than for the closed one (solid line). Similarly, figure~\ref{PBG3} shows an improvement in the algorithm performance for certain values of $\Delta_L$. This
results suggests that, at least within the two level approximation, a controlled
coupling with a certain environment may improve the performance of the quantum search algorithm.  Figures~\ref{PBG2} and~\ref{PBG4} show the result for the same parameter regimes as in~\ref{PBG1} and~\ref{PBG3} respectively, but just considering in
the equations the real part of the dissipation rates. Just as in the
case of the thermal environment, we can see that results for real and complex rates differ considerably.  

Indeed, considering the full complex rates is particularly important for a structured environment like the one analysed here. When the system frequency is within the gap, the corresponding rate is, at long times, a purely imaginary quantity. Hence, by only considering the real part of the rates one would arrive to the wrong conclusion that the environment has no effect at all in the system. However, it has been known for a long time that even when the system resonant frequency is within a gap of the spectral density, the coupling with the environment has important consequences in its dynamics. Among other things, a so-called photon-atom bound state is formed \cite{WJ03}, in which the energy is coherently interchanged between system and environment. This particular state, and more generally the dynamics of the system with a resonant frequency within the environment gap, can only be properly described when considering the imaginary part of the rates. 

In addition we note that, while in the thermal case the improvement of the algorithm performance observed for real rates can no longer be observed when considering the full complex rates, the opposite is observed here. Indeed, as noted above only when complex rates are included in the description, an increase in the final success probability is observed for certain parameter regimes. This can be seen again by comparing figures~\ref{PBG2} and~\ref{PBG4}, corresponding to real rates, with~\ref{PBG1} and~\ref{PBG3}.

From all this we again conclude how crucial it is
to consider the imaginary terms of the dissipation rates in order to
correctly describe the system dynamics.
%

\subsection{Application to a genuine two-level system}
In the former section we have seen within the two-level approximation,
how an adiabatic search algorithm can be improved by connecting the
system to a photonic crystal like environment as the one described in
\cite{deVega08}. We are aware that the applicability of this result to
a future implementation of the adiabatic version of the Grover's algorithm depends on how
good the two-level approximation is.  
The validity of the two-level approximation will be
analyzed in more detail in Sect.~\ref{sec:validity}.
Notice, however, that 
this caveat would not
apply to a genuine two-level system controllably coupled to a
dissipative environment.  
Therefore in this section we will study
if a \emph{real} two level system undergoing a more general adiabatic
evolution (not necessarily corresponding to the initial conditions of
Grover's algorithm) would also experience an improvement in the final
result under some condition. We recall that here an improvement means
that at the end of the adiabatic evolution, the population of the
ground state of the final Hamiltonian is larger than in the closed
system case.

To this order we consider the Hamiltonian (\ref{genH2}) but with
$n=1$, corresponding to a single site in the lattice. This model can
be realized by considering a lattice with a filling factor
sufficiently low so that the sparse atoms do not interact with each
other. Hence, in a real setup we would have several independent
copies of atoms at single sites. Instead of the adiabatic search algorithm, which
should be realized with $N$ sites, let us consider an adiabatic
process described by the Hamiltonian (\ref{Grov1}), but with $|m\rangle=|1\rangle$ and 
$|\psi_0\rangle=a_0|0\rangle+b_0|1\rangle$ with arbitrary coefficients
$a_0$ and $b_0=\sqrt{1-a_0^2}$.
\begin{figure}[ht]
\centerline{\includegraphics[width=0.5\textwidth]{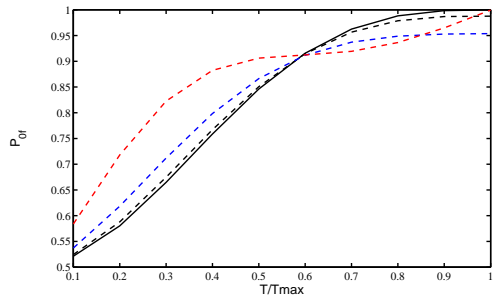}}
\caption{Final success probability (population of the ground state of the final Hamiltonian) for a genuine two level system undergoing adiabatic evolution, considering different adiabatic evolution times $T$, with $T_{\mmax}=10$ (the approximate time in which the closed system reaches a success probability equal to one). The solid line corresponds to the closed system, while black, green and red discontinuous lines correspond to the results for $\eta=0.01,0.05,0.2$ respectively. Other parameters are chosen as $\omega_0=0.5$, $\Delta_L=0.5$, and $a_0=\sqrt{0.5}$.}
\label{arb}
\end{figure}
We observe in Figure \ref{arb} that, for certain parameter regimes
the final probability of success for the open system is larger than
for the closed system. Similar results have been observed for other
choices of the coefficient $a_0$. Hence, we conclude that the
improvement of the adiabatic evolution for the system coupled with the
environment is a robust effect in the two-level system, 
not linked to the particular choice of initial state in the search algorithm.
Although in this section we have studied in particular a system subject to a photonic crystal-like environment, we believe that similar results could be encountered in other different scenarios, provided that the interaction between the system and the environment can be externally controlled. A simple example would be an atom in which the transition between the two relevant internal levels is produced by a two-photon process, mediated by a single laser followed by a spontaneous emission. This corresponds in quantum optics to a $\Lambda$ scheme, and would give rise to an effective Hamiltonian of the form (\ref{genH2}). Despite having a different dispersion relation, this Hamiltonian would describe a coupling with the environment that can be tuned through the laser Rabi frequency.

\section{Spectral density and the validity of the two level approximation}
\label{sec:validity}
Let us now qualitatively study the validity of the two-level
approximation. This approximation is suitable when the environment does not produce
significant transitions to other levels of the energy spectrum
different than $|0\rangle$ and $|1\rangle$, the ones involved in the
quantum search algorithm.  In general, the transition probability per unit time 
between two system levels $|i\rangle$ and $|j\rangle$ is approximately
given by the Fermi Golden Rule as $P_{ij}\approx J(\omega_{ij})\langle
i|H_\iint|j\rangle$. From this formula, we can already see that the spectral density, $J(\omega)$, is indeed a
very important quantity that determines, at each frequency, what is
the density of environmental states available, and how strong the
coupling is.  
Indeed, one may expect that the
two-level approximation applies better to parameter regimes such that the
spectral density is small for frequencies corresponding to transitions
to levels different from $|0\rangle$ and $|1\rangle$.

Notice that this can only be understood as a first qualitative
approach to the question of the validity of this approximation. Firstly, because an equally important factor to determine $P_{ij}$ is the magnitude of the 
transition amplitudes $\langle i|H_\iint|j\rangle$, and secondly, because $P_{ij}$ only accounts for the real part of the transition rates.

\begin{figure}[ht]
\centerline{\includegraphics[width=0.5\textwidth]{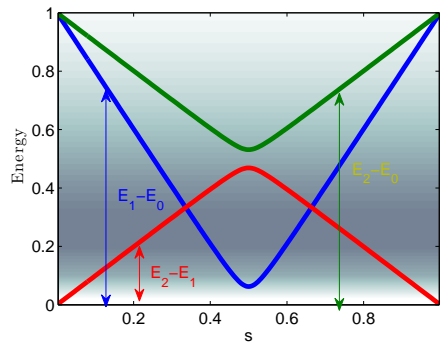}}
\caption{The energy gaps in the closed system as a function of the
  adiabatic parameter $s$: $E_1-E_0$ (blue), $E_2-E_1$ (red) and
  $E_2-E_0$ (green). For each value of energy, the background color
  indicates the magnitude of the spectral density for a thermal bath,
  where we have chosen the same parameters as in section~(\ref{thermal}), $s=1$ and $\omega_c=0.25$.}
\label{fig:GapJTherm}
\end{figure}

Let us illustrate this qualitative approach by considering the energy
vs. time plot in figure~\ref{fig:GapJTherm} for the thermal bath.  All
energy differences $E_i-E_j$ in the closed system are represented in
the plot as solid lines. The background shows how the spectral density
varies for different energies, with darker color representing a larger
value of $J$.  Intuitively we expect that a particular transition
$i\rightarrow j$ becomes more important when the corresponding energy
gap coincides with large values of $J(\omega)$.

%
In order to compute the spectral function for the photonic crystal-like environment, let us consider the calculus of the correlation function of the system as defined in (\ref{Gn0}),
\bea
g(t)=\frac{4\pi\Omega_L^2 X_0^3}{\pi^{3/2}}\int_0^\infty dk k^2 e^{-i (\frac{k^2}{2m}-\Delta_L)t}e^{-{X_0^2k^2}}.
\label{momento}
\eea
Here, both $g_k$ and $\omega_{\bf k}$ only depend on the modulus of the wave vector. Using this, the angular part of the integral has been performed (giving rise to a factor $4\pi$) and only the integral of the modulus is left. We can consider that the coupling parameter $g_{\bf k}$ does only depend on the modulus of the wave vector, because it has been assumed that the total wave vector of the Raman lasers ${\bf k}_L\approx 0$.
According to the definition (\ref{spectral}), the spectral function is the kernel of the integral (\ref{momento}) translated to the frequency space. Hence, it is necessary to use the dispersion relation $\omega(k)=\frac{k^2}{2m}+\omega_b$ in order to perform a change of variable in the integral (\ref{momento}), such that it becomes an integral in $\omega$,
\bea
g(t)=\eta\int_{\omega_b}^\infty d\omega \sqrt{2(\omega-\omega_b)}e^{-i (\omega-\omega_L)t}e^{-2\frac{\omega-\omega_b}{\omega_0}},
\label{frecuencia}
\eea
with $\eta=\frac{8\pi^{1/2}\Omega_L^2}{\omega_0^{3/2}}$. Considering now an energy shift of the form $\hat{\omega}=\omega-\omega_L$, the former integral is just
\bea
g(t)=\eta\int_{\Delta_L}^\infty d\omega \sqrt{2(\omega-\Delta_L)}e^{-i \omega t}e^{-2\frac{\omega-\Delta_L}{\omega_0}}.
\label{frecuencia2}
\eea
Since in this case the temperature of the reservoir is zero, it is straightforward to see that the spectral density has the form  $J(\omega)=\eta\sqrt{2(\omega-\Delta_L)}e^{-2\frac{\omega-\Delta_L}{\omega_0}}$. 

\begin{figure}[ht]
\centerline{\includegraphics[width=0.5\textwidth]{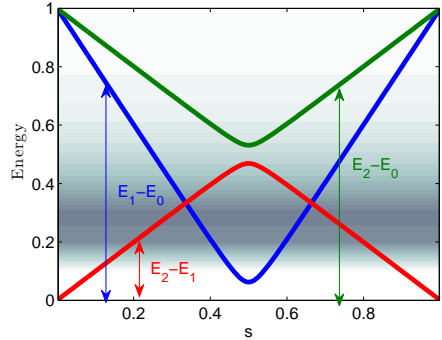}}
\caption{The energy gaps in the closed system as a function of the
  adiabatic parameter $s$: $E_1-E_0$ (blue), $E_2-E_1$ (red) and
  $E_2-E_0$ (green). The background indicates the magnitude of the
  spectral density for the photonic crystal-like environment, with
  $\Delta_L=0.28$ and $\omega_0=0.25$ as typical parameters used in section~(\ref{structured}).  We expect that the dominant
  transitions induced by the couplig to the environment are
  $0\leftrightarrow 1$. The transition $1\leftrightarrow 2$ may also
  contribute, but it will be suppressed during the first part of the
  algorithm because the $E_1$ level is initially not populated. It
  will again be important after $s\approx0.65$. With these parameters,
  moreover, the minimum gap, around $s=0.5$, is \emph{decoupled} from
  the frequencies of the bath, what can be expected to bring some
  protection to the most sensitive part of the algorithm.}
\label{fig:GapJPBG}
\end{figure}

Like in the thermal case, figure~\ref{fig:GapJPBG} represents all energy differences $E_i-E_j$ for 
the closed system in solid lines, compared to the spectral density in the background. From a naive interpretation of the picture, one would consider that, since the minimum gap $E_1-E_0$ corresponds to an energy within the environment gap, where the spectral density is zero, the effects of the environment are somehow minimized. However, as it is shown in figure ~\ref{PBG1}, for the same $\Delta_L=0.28$, the performance of the algorithm is even improved with respect to that of the closed system. Hence, the environment gap does not protect the algorithm from the interaction, but rather it produces some positive effects on its performance.
As already mentioned, another important factor that determines the transition
probability to other levels $|i\rangle$ different to the ones
considered in the two level approximation ($|0\rangle$ and $|1\rangle$)
is the amplitude of the transition elements $\langle
i|H_\iint|0\rangle$, $\langle i|H_\iint|1\rangle$, and in general
$\langle i|H_\iint|j\rangle$, with $H_\iint$ given by
(\ref{genH2}). However, this analysis is out of the scope of the
present paper, and will be made elsewere.
\section{Conclusions}

We have studied the performance of the adiabatic quantum search in a
dissipative environment. Our derivation of the Bloch-Redfield equation
allows for a more complete account of the effects of the 
complex dissipation rates than previous studies. 
In particular, we have shown that neglecting the imaginary parts of such rates give rise to a completely different dynamic than when considering the full complex rates. Hence, conclusions about whether there is an improvement of the performance of the adiabatic algorithm within an open system, cannot be extracted by a partial analysis that only accounts for the real part of the dissipation rates.


Particularly, we have analyzed the effect of the dissipation on the
adiabatic version of Grover's algorithm for two different settings,
namely a thermal bath and a controllable environment with tunable
parameters. While for a thermal environment no improvement is observed
in the performance of the algorithm, coupling the system to a
structured environment gives rise to a final success probability
that for certain parameter regimes is higher than in the closed system
case. Hence, we have found an example in which the performance of the
search can be improved with respect to the closed system, by tuning
the bath parameters appropriately.

Our study, valid for the quantum search in the framework of the
two-level approximation, can be extended to the adiabatic evolution of 
a genuine two-dimensional system.

We acknowledge J. I. Cirac, S. F. Huelga,
  M. B. Plenio for support, and A. M. S. Amin for interesting
  discussions. We also thank 'Centro de ciencias de Benasque' for its hospitality. This work has been supported in part by the Spanish
  Ministerio de Educaci\'on y Ciencia through Projects
  AYA2007-67626-C03-01 and FPA2008-03373, by the 'Generalitat Valenciana' grant PROMETEO/2009/128, by the DFG Cluster of
  Excellence MAP, the SFB 631, and the DFG Forschergruppe 635, and by
  the EU (FP7/2007-2013) under grant agreement nr. 247687 (Integrating
  Project AQUTE).  

\section{Appendix A}
\label{AppendixA}
In this appendix we proceed to derive equation (\ref{red2}) in the paper.
The total density matrix can be written in the form
\bea
\rho_\ttot (t)=\rho_S(t)\otimes \rho_B(t) + \rho_\ccor ,
\label{totald}
\eea 
where $\rho_S = Tr_B( \rho_\ttot (t))$ and $\rho_B = Tr_S( \rho_\ttot (t))$ are
the reduced density matrices of the system and the bath, respectively, 
and  $\rho_\ccor $ describes the correlations existing between
system and environment. We assume that in general we are dealing
with a thermal bath, so that $ \rho_B=e^{-\beta
  H_B}$, with $\beta=1/(K_B T)$, $K_B$ the Boltzmann constant and
$T$ the temperature of the reservoir.  We consider the master
equation for the reduced density operator in the interaction picture with
respect to both the system and the environment, $\rho_{SI}={\mathcal
  U}^\dagger(t) \rho_S {\mathcal U}(t)$,  
where $\mathcal{U}(t)$ is given by the
time-ordered exponential ${\mathcal  U}(t)=Te^{-i\int_0^t H_S(\tau))d\tau}$.
Considering that $\frac{d{\mathcal
    U}^\dagger(t)}{dt}=iH_S(t){\mathcal U}^\dagger(t)$, the evolution
equation up to second order in $g=\|H_\iint \|/\|H_S+H_B\|$ is given
by \cite{amin2} \bea \dot{\rho}_{SI}(t)=-\int_0^t d\tau Tr_B[H_I
  (t),[H_I (\tau),\rho_{SI}(t)\otimes\rho_B]].
\label{interact}
\eea
Here the terms $\sim Tr_B (\rho_B H_\iint ^t) $ have been neglected, an assumption that is valid for most types of environments and couplings. Particularly, it is valid for our present case of a system lineraly coupled to an environment of harmonic oscillators.
In addition, it has also been assumed that $\rho(t)=\rho_S(t)\otimes\rho_B$, and since $\tau_C\ll 1/\Gamma$, we have also assumed that $\rho_{SI}(\tau)\approx \rho_{SI}(t)$ in the integral.
On the other hand, the quantity $H_I (t)={\mathcal U}^\dagger(t) H^t_\iint {\mathcal U}(t)$ with $H^t_\iint=e^{iH_B t}H_{\iint}e^{-iH_B t}$. Going back to the original picture, the evolution equation of $\rho_S$ is then given by
\bea
\frac{d\rho_S}{dt}&=&-i[H_S(t),\rho_S]\nonumber\\
&-&\int_0^t d\tau {\mathcal U}(t) Tr_B[H_I (t),[H_I (\tau),\rho_{SI}(t)\otimes\rho_B]]{\mathcal U}^\dagger(t).
\eea
The latter equation can also be expressed as
\bea
\frac{d\rho_S}{dt}=-i[H_S(t),\rho_S]-\int_0^t d\tau Tr_B\left[H^t_\iint,[H_I (\tau,-t),\rho_{S}(t)\otimes\rho_B]\right],
\label{red1}
\eea
where $H_I (\tau,-t)={\mathcal U}(t){\mathcal U}^\dagger(\tau)H^\tau_\iint {\mathcal U}(\tau){\mathcal U}^\dagger(t)$.
Expanding the commutator on the right hand side of the last equation, we find the following terms,
\bea
&Tr_B&\left([H^t_\iint,[H_I (\tau,-t),\rho_{S}(t)\rho_B]]\right)=Tr_B\big(\{H^t_\iint H_I(\tau,-t),\rho_S\otimes\rho_B\}\nonumber\\
&-&H^t_\iint \rho_S\otimes\rho_B H_I(\tau,-t)-H_I(\tau,-t)\rho_S\otimes\rho_B H^t_\iint\big).
\label{tr}
\eea
Let us now consider our case, in which the interaction Hamiltonian can be written as  $H_\iint =A\otimes B$, where $A=\sigma_z$ and $B=\sum_\lambda g_\lambda(b_\lambda^\dagger+b_\lambda)$. We can make further simplifications in equation (\ref{tr})
\bea
&Tr_B&\left([H^t_\iint,[H_I (\tau,-t),\rho_{S}(t)\otimes\rho_B]]\right)\nonumber\\
&=&g(t-\tau)\left(AA(\tau,-t)\rho_S-A(\tau,-t)\rho_S A\right)\nonumber\\
&+&g(\tau-t)\left(\rho_S A(\tau,-t)A-A\rho_S A(\tau,-t)\right),
\label{con}
\eea
where $A(\tau-t)={\mathcal U}(t){\mathcal U}^\dagger(\tau)A {\mathcal U}(\tau){\mathcal U}^\dagger(t)$, and $g(t-\tau)=Tr_B[B^t B^\tau \rho_B]$, $g(\tau-t)=Tr_B[B^\tau B^t \rho_B]$, having the form
\begin{eqnarray}
g(t)= \sum_\lambda g_\lambda^2 [\coth{\left(\frac{\beta\omega_\lambda}{2}\right)}\cos{(\omega_\lambda t)}-i\sin{(\omega_\lambda t)}].
\label{correlh2}
\end{eqnarray}
Here, we have taken into account that, for a thermal bath, 
\bea
Tr_B(\rho_B b_\lambda^\dagger b_{\lambda'})=\delta_{\lambda,\lambda'}N(\omega_\lambda),\nonumber\\
Tr_B(\rho_B b_\lambda b^\dagger_{\lambda'})=1+\delta_{\lambda,\lambda'}N(\omega_\lambda),
\eea
with $N(\omega_\lambda)=1/(e^{\omega_\lambda \beta}-1)$ the number of excitations with frequency $\omega_\lambda$. Note that according to equation (\ref{correlh}), $g(-t)=g^*(t)$.
Considering that, and inserting (\ref{con}) in (\ref{red1}) we find
\bea
\frac{d\rho_S}{dt}&=&-i[H_S(t),\rho_S]-\int_0^t d\tau g(t-\tau)\left(AA(\tau,-t)\rho_S-A(\tau,-t)\rho_S A\right)\nonumber\\
&-&\int_0^t d\tau g^*(t-\tau)\left(\rho_S A(\tau,-t)A-A\rho_S A(\tau,-t)\right).
\label{red3}
\eea

\bibliography{adiabatic6}
\bibliographystyle{unsrt}
\end{document}